\documentclass{article}

\usepackage{times}
\usepackage{amssymb,amsmath,amsthm} 
\usepackage{mathrsfs} 
\usepackage{epsfig}
\usepackage{color}

\newcommand{\ie}{\emph{i.e.}~} 
\newcommand{\cf}{\emph{cf.}~}

\def\A{\mathcal A}
\def\B{\mathscr B}
\def\BB{\mathcal B}
\def\C{\mathbb C}

\def\EE{\mathcal E}
\def\F{\mathbb F}
\def\FF{\mathcal F}
\def\G{\mathcal G}
\def\H{\mathcal H}
\def\K{\mathcal K}

\def\M{\mathcal M}
\def\N{\mathbb N}

\def\R{\mathbb R}

\def\Z{\mathbb Z}

\def\<{\left\langle} 
\def\>{\right\rangle} 
\def\({\left(} 
\def\){\right)} 
\def\[{\left[} 
\def\]{\right]} 
 
\def\e{\mathop{\mathrm{e}}\nolimits} 
\def\d{\mathrm{d}} 
\def\supp{\mathop{\mathrm{supp}}\nolimits}

\def\im{\mathop{\mathsf{Im}}\nolimits}
\def\re{\mathop{\mathsf{Re}}\nolimits}
\def\ind{\mathop{\mathrm{ind}}\nolimits}

\newtheorem{Theorem}{Theorem}[section] 
\newtheorem{Remark}[Theorem]{Remark} 
\newtheorem{Lemma}[Theorem]{Lemma} 
 
\newtheorem{Corollary}[Theorem]{Corollary} 
\newtheorem{Proposition}[Theorem]{Proposition} 
\newtheorem{Definition}[Theorem]{Definition} 

\begin{document}


\title{\Large\textbf{Spectral analysis for adjacency operators on graphs}}
   
\author{M. M\u antoiu$^1$, S. Richard$^2$ and R. Tiedra de Aldecoa$^3$} 
\date{\small}
\maketitle
\vspace{-1cm}

\begin{quote}
\emph{
\begin{itemize}
\item[$^1$] Institute of Mathematics ``Simion Stoilow'' of the Romanian Academy,
P.\,O. Box 1-764, 014700 Bucharest, Romania
\item[$^2$] Institut Camille Jordan, Universit\'e Claude Bernard Lyon 1,
43 boulevard du 11 novembre 1918, 69622 Villeurbanne cedex, France
\item[$^3$] D\'epartement de math\'ematiques, Universit\'e de Paris XI, 91405 Orsay Cedex,
France 
\item[] \emph{E-mails:} Marius.Mantoiu@imar.ro, srichard@math.univ-lyon1.fr,\\
	    rafael.tiedra@math.u-psud.fr 
\end{itemize}
  }
\end{quote}


\begin{abstract} 
We put into evidence graphs with adjacency operator whose singular subspace is prescribed by
the kernel of an auxiliary operator. In particular, for a family of graphs called admissible,
the singular continuous spectrum is absent and there is at most an eigenvalue located at the
origin. Among other examples, the one-dimensional XY model of solid-state physics is
covered. The proofs rely on commutators methods.
\end{abstract}

\textbf{Key words and phrases:} graph, adjacency operator, positive commutator, singular spectrum.

\textbf{2000 Mathematics Subject Classification:} 05C20, 47A10, 47B39.

\section{Introduction}
\setcounter{equation}{0} 

Let $(X,\sim)$ be a graph. We write $x\sim y$ whenever the vertices (points) $x$ and $y$ of $X$
are connected. For simplicity, we do not allow multiple edges or loops. In the Hilbert space
$\H:=\ell^2(X)$ we consider the \emph{adjacency operator}
\begin{equation*}
(Hf)(x):=\sum_{y\sim x}f(y),\quad f\in\H,~x\in X.
\end{equation*}
We denote by $\deg(x):=\#\{y\in X\:\!:\:\!y\sim x\}$ the degree of the vertex $x$. Under the
assumption that $\deg(X):=\sup_{x\in X}\deg(x)$ is finite, $H$ is a bounded selfadjoint
operator in $\H$. We are interested in the nature of its spectral measure. Useful sources
concerning operators acting on graphs are \cite{Bang/Gutin,Mohar,Mohar/Woess}, see also the
references therein.

Rather few adjacency operators on graphs are known to have purely absolutely continuous
spectrum. This occurs for the lattice $\mathbb Z^n$ and for homogeneous trees. These and
several other examples are presented briefly in \cite{Mohar/Woess}. Adjacency operators may
also have non-void singular spectrum. In \cite{Simon96} the author exhibits families of
ladder-type graphs for which the existence of singular continuous spectrum is generic.
Percolation graphs with highly probable dense pure point spectrum are presented in
\cite{Veselic05}, see also \cite{Kirk/Egg} and \cite{CCFST} for earlier works.

In the sequel we use commutator methods to study the nature of the spectrum of adjacency
operators. Mourre theory \cite{Mourre81,ABG}, already applied to operators on trees
\cite{Allard/Froese,Georgescu/Golenia}, may be a well-fitted tool, but it is not easy to use
it in non-trivial situations. We use a simpler commutator method, introduced in
\cite{BKM,Boutet/Mantoiu} and called ``the method of the weakly conjugate operator''.
It is an unbounded version of the Kato-Putnam theorem \cite{RS}, which will be presented
briefly in Section \ref{sec1}.

The method of the weakly conjugate operator provides estimates on the behaviour of the
resolvent $(H-z)^{-1}$ when $z$ approaches the spectrum of $H$. These estimates are global, 
\ie uniform in $\re(z)$. They imply precise spectral properties for $H$. For the convenience
of the reader, we are going to state now spectral results only in the particular case of
``admissible graphs'' introduced in Section \ref{secadmis}. The general results, including
boundary estimates for the resolvent and perturbations, are stated in Section \ref{sec3} and
proved in Section \ref{proofmain}.

The notion of admissibility requires (among other things) the graph to be directed. Thus the
family of neighbours $N(x):=\{y\in X\:\!:\:\!y\sim x\}$ is divided into two disjoint sets
$N^-(x)$ (fathers) and $N^+(x)$ (sons), $N(x)=N^-(x)\sqcup N^+(x)$. We write $y<x$ if
$y\in N^-(x)$ and $x<y$ if $y\in N^+(x)$. On drawings, we set an arrow from $y$ to $x$
($x\leftarrow y$) if $x<y$, and say that the edge from $y$ to $x$ is positively oriented.

We assume that the subjacent directed graph, from now on denoted by $(X,<)$, is \emph{admissible} with 
respect to these decompositions, \ie (i) it admits a position function and (ii) it is uniform. A
\emph{position function} is a function $\Phi:X\to\Z$ such that $\Phi(y)+1=\Phi(x)$ whenever
$y<x$.  It is easy to see that it exists if and only if all paths between two points have
the same index (which is the difference between the number of positively and negatively
oriented edges). Position functions and the number operator from
\cite[Section 2]{Georgescu/Golenia} present some common features. The directed graph $(X,<)$ is called
\emph{uniform} if for any $x,y\in X$ the number $\#\[N^-(x)\cap N^-(y)\]$ of common fathers
of $x$ and $y$ equals the number $\#\[N^+(x)\cap N^+(y)\]$ of common sons of $x$ and $y$.
Thus the admissibility of a directed graph is an explicit property that can be checked
directly, without making any choice. The graph $(X,\sim)$ is \emph{admissible} if there exists 
an admissible directed graph subjacent to it.

\begin{Theorem}\label{first}
The adjacency operator of an admissible graph $(X,\sim)$ is purely absolutely continuous,
except at the origin, where it may have an eigenvalue with eigenspace
\begin{equation}\label{ma_belle_courgette}
\ker(H)=\big\{f\in\H\:\!:\:\!\textstyle\sum_{y<x}f(y)=0=\sum_{y>x}f(y)
~\textrm{ for each}~x\in X\big\}.
\end{equation}
\end{Theorem}

Theorem \ref{doi}, which is more general, relies on the existence of a \emph{function
adapted to the graph}, a concept generalising that of a position function. Examples of
graphs, both admissible and non-admissible, are presented in Section \ref{someother}.
In Section \ref{Dp} we treat $D$-products of graphs.

Our initial motivation in studying the nature of the spectrum of operators on graphs comes
from spin models on lattices. We refer to \cite{DMT} for some results on the essential
spectrum and localization properties for the one-dimensional Heisenberg model and for more
general Toeplitz-like operators. In the final section of the present article we show that the
one-dimensional $XY$ model has a purely absolutely continuous spectrum, except maybe at the
origin (see Corollary \ref{presque} and Remark \ref{pourquoipresque}). This result (which we
have not found in the literature) is shown using a very direct approach, involving
only pure operator methods. The more complex one-dimensional Heisenberg model is
analysed in detail in \cite{Babbitt/Thomas77(II)} and \cite{Babbitt/Thomas77(IV)}.

\section{The method of the weakly conjugate operator}\label{sec1}
\setcounter{equation}{0} 

In this section we recall the basic characteristics of the method of the weakly conjugate
operator. It was introduced and applied to partial differential operators in
\cite{BKM,Boutet/Mantoiu}. Several developments and applications may be found in
\cite{Iftimovici/Mantoiu,Mantoiu/Pascu,Mantoiu/Richard,Richard06}. The method works for
unbounded operators, but for our purposes it will be enough to assume $H$ bounded.

We start by introducing some notations. The symbol $\H$ stands for a Hilbert space with scalar
product $\<\:\!\cdot\:\!,\:\!\cdot\:\!\>$ and norm $\|\cdot\|$. Given two Hilbert spaces
$\H_1$ and $\H_2$, we denote by $\B(\H_1,\H_2)$ the set of bounded operators from $\H_1$ to
$\H_2$, and put $\B(\H):=\B(\H,\H)$.  We assume that $\H$ is endowed with a strongly continuous
unitary group $\{W_t\}_{t\in\R}$. Its selfadjoint generator is denoted by $A$ and has domain
$D(A)$. In most of the applications $A$ is unbounded.

\begin{Definition}\label{defC1}
A bounded selfadjoint operator $H$ in $\H$ belongs to $C^1(A;\H)$ if one of the following
equivalent condition is satisfied:
\begin{enumerate} 
\item[(i)] the map $\R\ni t\mapsto W_{-t}HW_t\in\B(\H)$ is strongly differentiable,
\item[(ii)] the sesquilinear form
\begin{equation*}
D(A)\times D(A)\ni(f,g)\mapsto i\<Hf,Ag\>-i\<Af,Hg\>\in\C
\end{equation*}
is continuous when $D(A)$ is endowed with the topology of $\H$.
\end{enumerate}
\end{Definition}

We denote by $B$ the strong derivative in (i), or equivalently the bounded selfadjoint
operator associated with the extension of the form in (ii). The operator $B$ provides a
rigorous meaning  to the commutator $i[H,A]$. We shall write $B>0$ if $B$ is positive and
injective, namely if $\<f,Bf\>>0$ for all $f\in\H\setminus\{0\}$.

\begin{Definition}
The operator $A$ is \emph{weakly conjugate to} the bounded selfadjoint operator $H$ if
$H\in C^1(A;\H)$ and $B\equiv i[H,A]>0$.
\end{Definition}

For $B>0$ let us consider the completion $\BB$ of $\H$ with respect to the norm 
$\|f\|_\BB:=\<f,Bf\>^{1/2}$. The adjoint space $\BB^*$ of $\BB$ can be identified with the
completion of $B\H$ with respect to the norm $\|g\|_{\BB^*}:=\<g,B^{-1}g\>^{1/2}$. One has
then the continuous dense embeddings $\BB^*\hookrightarrow\H\hookrightarrow\BB$, and $B$
extends to an isometric operator from $\BB$ to $\BB^*$. Due to these embeddings it makes
sense to assume that $\{W_t\}_{t\in \R}$ restricts to a $C_0$-group in $\BB^*$, or
equivalently that it extends to a $C_0$-group in $\BB$. Under this assumption (tacitly
assumed in the sequel) we keep the same notation for these $C_0$-groups. The domain of the
generator of the $C_0$-group in $\BB$ (resp. $\BB^*$) endowed with the graph norm is denoted
by $D(A,\BB)$ (resp.~$D(A,\BB^*)$). In analogy with Definition \ref{defC1} the requirement
$B\in C^1(A;\BB,\BB^*)$ means that the map $\R\ni t\mapsto W_{-t}BW_t\in\B(\BB,\BB^*)$ is
strongly differentiable, or equivalently that the sesquilinear form
\begin{equation*}
D(A,\BB)\times D(A,\BB)\ni(f,g)\mapsto i\<f,BAg\>-i\<Af,Bg\>\in\C
\end{equation*}
is continuous when $D(A,\BB)$ is endowed with the topology of $\BB$. Here,
$\<\:\!\cdot\:\!,\:\!\cdot\:\!\>$ denotes the duality between $\BB$ and $\BB^*$. Finally let
$\EE$ be the Banach space $\big(D(A,\BB^*), \BB^*\big)_{1/2,1}$ defined by real interpolation
(see for example \cite[Proposition 2.7.3]{ABG}). One has then the natural continuous embeddings 
$\B(\H)\subset\B(\BB^*,\BB)\subset\B(\EE,\EE^*)$ and the following results
\cite[Theorem 2.1]{Boutet/Mantoiu}:

\begin{Theorem}\label{thmabstract}
Assume that $A$ is weakly conjugate to $H$ and that $B\equiv i[H,A]$ belongs to
$C^1(A;\BB,\BB^*)$. Then there exists a constant $\textsc c>0$ such that
\begin{equation}\label{liap}
\left|\<f,(H-\lambda\mp i\mu)^{-1}f\>\right|\leq\textsc c\|f\|_\EE^2
\end{equation}
for all $\lambda\in\R$, $\mu>0$ and $f\in\EE$. In particular the spectrum of $H$ is purely
absolutely continuous.
\end{Theorem}

For readers not accustomed with real interpolation or with the results of \cite{ABG}, we
mention that one can replace $\|f\|_\EE$ by $\|f\|_{D\(A,\BB^*\)}$ in Formula (\ref{liap}),
loosing part of its strength. In the applications it may even be useful to consider smaller,
but more explicit, Banach spaces $\FF$ continuously and densely embedded in $D(A,\BB^*)$.
In such a setting we state a corollary of Theorem \ref{thmabstract}, which follows by applying
the theory of smooth operators \cite{RS,BKM}. The adjoint space of $\FF$ is denoted by
$\FF^*$.

\begin{Corollary}\label{JabaTheHut}
\begin{enumerate}
\item[(a)] If $\,T$ belongs to $\B\(\FF^*,\H\)$, then $T$ is an $H$-smooth operator.
\item[(b)] Let $U$ be a bounded selfadjoint operator in $\H$ such that $|U|^{1/2}$ extends to
an element of  $\B\(\FF^*,\H\)$. For $\gamma \in \R$, let $H_\gamma:=H+\gamma U$. Then there
exists $\gamma_0>0$ such that for $\gamma \in (-\gamma_0,\gamma_0)$, $H_\gamma:=H+\gamma U$ is
purely absolutely continuous and unitarily equivalent to $H$ through the wave operators
$\Omega_\gamma^\pm :=\emph{s-}\lim_{t\to\pm \infty}\e^{itH_\gamma}\e^{-itH}$.
\end{enumerate}
\end{Corollary}

\section{Statement of the main result}\label{sec3}
\setcounter{equation}{0} 

Some preliminaries on graphs could be convenient, since notations and conventions do not seem
commonly accepted in graph theory.

A {\it graph} is a couple $(X,\sim)$ formed of a non-void countable set $X$ and a symmetric
relation $\sim$ on $X$ such that $x\sim y$ implies $x\ne y$. The points $x\in X$ are called
\emph{vertices} and couples $(x,y)\in X\times X$ such that $x\sim y$ are called \emph{edges}.
So, for simplicity, multiple edges and loops are forbidden in our definition of a graph.
Occasionally $(X,\sim)$ is said to be a \emph{simple} graph.

For any $x\in X$ we denote by $N(x):=\{y\in X\:\!:\:\!y\sim x\}$ the set of \emph{neighbours}
of $x$. We write $\deg(x):=\#N(x)$ for the \emph{degree} or \emph{valence} of the vertex $x$
and $\deg(X):=\sup_{x\in X}\deg(x)$ for the degree of the graph. We also suppose that $(X,\sim)$
is \emph{uniformly locally finite}, \ie that $\deg(X)<\infty$. When the function
$x\mapsto\deg(x)$ is constant, we say that the graph is {\it regular}.

A \emph{path} from $x$ to $y$ is a sequence $p=(x_0,x_1,\dots,x_n)$ of elements of $X$,
usually denoted by $x_0x_1\dots x_n$, such that $x_0=x$, $x_n=y$ and $x_{j-1}\sim x_j$ for
each $j\in\{1,\dots,n\}$. The \emph{length} of the path $p$ is the number $n$. If $x_0=x_n$ we
say that the path is \emph{closed}. A graph is \emph{connected} if there exists a path
connecting any two vertices $x$ and $y$. On any connected graph $(X,\sim)$ one may define
the \emph{distance function} $d$ as follows: $d(x,x):=0$ and $d(x,y)$ is equal to the length
of the shortest path from $x$ to $y$ if $x\ne y$.

Throughout the paper we restrict ourselves tacitly to graphs $(X,\sim)$ which are simple,
infinite countable and uniformly locally finite. Given such a graph we consider the
\emph{adjacency operator} $H$ acting in the Hilbert space $\H:=\ell^2(X)$ as
\begin{equation*}
(Hf)(x):=\sum_{y\sim x}f(y),\quad f\in\H,~x\in X.
\end{equation*}
Due to \cite[Theorem 3.1]{Mohar/Woess}, $H$ is a bounded selfadjoint operator with
$\|H\|\le\deg(X)$ and spectrum $\sigma(H)\subset[-\deg(X),\deg(X)]$. If $(X,\sim)$ is not
connected, $H$ can be written as a direct sum in an obvious manner and each component can be
treated separately. Most of the time $(X,\sim)$ will be assumed to be connected.

For further use, we also sketch some properties of a larger class of operators. Any element
of $\B[\ell^2(X)]$ is an ``integral" operator of the form
$\(I_af\)(x)=\sum_{y\in X}a(x,y)f(y)$ for some matrix $a\equiv\{a(x,y)\}_{x,y\in X}$. Formally
$I_a$ is symmetric if and only if $a$ is symmetric, \ie $\overline{a(x,y)}=a(y,x)$, and
$I_a,I_b$ satisfy the multiplication rule $I_a I_b=I_{a\circ b}$ with
$(a\circ b)(x,y):=\sum_{z\in X}a(x,z)b(z,y)$. A bound on the norm of $I_a$ is given by the
relation
\begin{equation}\label{Schur}
\|I_a\|\le\max\left\{\textstyle\sup_{x\in X}\sum_{y\in X}|a(x,y)|\,
,\,\sup_{y\in X}\sum_{x\in X}|a(x,y)|\right\}.
\end{equation}
In the sequel we shall encounter only matrices $a\in\ell^\infty(X\times X)$ such that there
exists a positive integer $k$ with
$\max\left\{\#[\supp a(x,\:\!\cdot\:\!)]\:\!,\:\!\#[\supp a(\:\!\cdot\:\!,x)]\right\}\leq k$
for all $x\in X$. Then an easy calculation using Formula \eqref{Schur} gives
$\|I_a\|\leq k\,\|a\|_\infty$. In particular we call \emph{local} an operator $I_a$ for which
$a(x,y)\neq 0$ only if $x\sim y$. In this case, if $a$ is symmetric and bounded, then $I_a$ is
selfadjoint and bounded, with $\|I_a\|\leq \deg(X)\,\|a\|_\infty$.

The methods of this article apply to the latter class of operators (commutator calculations
involve operators $I_a$ which are not local, but bounded since they satisfy $a(x,y)=0$ if
$d(x,y)\geq 3$). However we refrained from treating more general objects than adjacency
operators for simplicity and because we have nothing remarkable to say about the general
case.

We now introduce the key concept. Sums over the empty set are zero by convention.

\begin{Definition}\label{adaptat}
A function $\Phi: X\to\R$ is \emph{semi-adapted} to the graph $(X,\sim)$ if
\begin{enumerate}
\item[(i)] there exists $\textsc c\ge0$ such that $|\Phi(x)-\Phi(y)|\le\textsc c\,$ for all
$x,y\in X$ with $x\sim y$,
\item[(ii)] for any $x,y\in X$ one has
\begin{equation}\label{semi}
\sum_{z\in N(x)\cap N(y)}[2\Phi(z)-\Phi(x)-\Phi(y)]=0.
\end{equation}
\end{enumerate}
If in addition for any $x,y\in X$ one has
\begin{equation}\label{full}
\sum_{z\in N(x)\cap N(y)}\[\Phi(z)-\Phi(x)\]\[\Phi(z)-\Phi(y)\]
\[2\Phi(z)-\Phi(x)-\Phi(y)\]=0,
\end{equation}
then $\Phi$ is \emph{adapted} to the graph $(X,\sim)$.
\end{Definition}

Let $M_Z(\Phi)$ be the mean of the function $\Phi$ over a finite subset $Z$ of $X$, \ie
$M_Z(\Phi):=(\#Z)^{-1}\sum_{z\in Z}\Phi(z)$. One may then rephrase Condition \eqref{semi}
as 
\begin{equation*}
M_{\{x,y\}}(\Phi)=M_{N(x)\cap N(y)}(\Phi)\quad\textrm{for any}~x,y\in X.
\end{equation*} 
In particular, if $x=y$, one simply has to check that
$\Phi(x)=\[\deg(x)\]^{-1}\sum_{y\sim x}\Phi(y)$ for all $x\in X$.

In order to formulate the main result we need a few more definitions. For a function $\Phi$  
semi-adapted to the graph $(X,\sim)$ we consider in $\H$ the operator $K$ given by
\begin{equation*}
(Kf)(x):=i\sum_{y\sim x}\[\Phi(y)-\Phi(x)\]f(y),\quad f\in\H,~x\in X.
\end{equation*}
The operator $K$ is selfadjoint and bounded due to the condition (i) of Definition
\ref{adaptat} and the discussion preceding it. It commutes with $H$, as a consequence of
Condition (\ref{semi}). We also decompose the Hilbert space $\H$ into the direct sum
$\H=\K\oplus\G$, where $\G$ is the closure of the range $K\H$ of $K$, thus the orthogonal
complement of the closed subspace
\begin{equation*}
\K:=\ker(K)=\left\{f\in\H\:\!:\:\!\textstyle\sum_{y\in N(x)}\Phi(y)f(y)
=\Phi(x)\sum_{y\in N(x)}f(y)~\textrm{ for each}~x\in X\right\}.
\end{equation*}
It is easy to see that $H$ and $K$ are reduced by this decomposition. Their restrictions
$H_0$ and $K_0$ to the Hilbert space $\G$ are bounded selfadjoint operators. The proofs of
the following results are given in the next section.

\begin{Theorem}\label{unu}
Assume that $\Phi$ is a function semi-adapted to the graph $(X,\sim)$. Then $H_0$ has no
point spectrum.
\end{Theorem}

In order to state a limiting absorption principle for $H_0$ in the presence of an adapted
function, we introduce an auxiliary Banach space. We denote by $\FF$ the completion of
$K\H\cap D(\Phi)$ with respect to the norm $\|f\|_\FF:=\||K_0|^{-1}f\|+\|\Phi f\|$ and we
write $\FF^*$ for the adjoint space of $\FF$. We shall prove subsequently the existence of
the continuous dense embeddings $\FF\hookrightarrow\G\hookrightarrow\FF^*$ and the following
result:

\begin{Theorem}\label{doi}
Let $\Phi$ be a function adapted to the graph $(X,\sim)$. Then
\begin{enumerate}
\item[(a)] There exists a constant $\textsc c>0$ such that
$\left|\<f,(H_0-\lambda\mp i\mu)^{-1}f\>\right|\leq\textsc c\,\|f\|_\FF^2\,$ for all
$\lambda\in\R$, $\mu>0$ and $f\in\FF$.
\item[(b)] The operator $H_0$ has a purely absolutely continuous spectrum.
\end{enumerate}
\end{Theorem}

In the next section we introduce a larger space $\EE$ obtained by real interpolation. The
limiting absorption principle is then obtained between the space $\EE$ and its adjoint
$\EE^*$. Of course, everything is trivial when $\K=\H$. This happens if and only if $\Phi$ is
a constant function (obviously adapted to any graph). We shall avoid this trivial case in the
examples. In many situations the subspace $\K$ can be calculated explicitly. On the other hand, 
if several adapted functions exist, one may use this to enlarge the space $\G$ on which $H$ is 
proved to be purely absolutely continuous.

The following result on the stability of the nature of the spectrum of $H_0$ under small
perturbations is a direct consequence of Corollary \ref{JabaTheHut}.

\begin{Corollary}
Let $U_0$ be a bounded selfadjoint operator in $\G$ such that $|U_0|^{1/2}$ extends to an
element of $\B(\FF^*,\G)$. Then, for $|\gamma|$ small enough, the operator $H_0+\gamma U_0$ is
purely absolutely continuous and is unitarily equivalent to $H_0$ through the wave operators.
\end{Corollary}

\section{Proof of the main result}\label{proofmain}
\setcounter{equation}{0} 

In this section we choose and fix a semi-adapted function $\Phi$. As a consequence of Condition
\eqref{semi}, one checks easily that the bounded selfadjoint operators $H$ and $K$ commute.
Aside $H$ and $K$ we also consider the operator $L$ in $\H$ given by
\begin{equation*}
(Lf)(x):=-\sum_{y\sim x}[\Phi(y)-\Phi(x)]^2f(y),\quad f\in\H,~x\in X.
\end{equation*}
Due to the discussion in Section \ref{sec3}, the  operator $L$ is selfadjoint and bounded.
Furthermore one may verify that $H$, $K$ and $L$ leave invariant the domain $D(\Phi)$ of the
operator of multiplication $\Phi$ and that one has on $D(\Phi)$ the relations
\begin{equation*}
K=i[H,\Phi],\qquad L=i[K,\Phi].
\end{equation*}
These relations imply that $H$ and $K$ belong to $C^1(\Phi;\H)$ (see Definition \ref{defC1}).
If in addition $\Phi$ is adapted to the graph, formula \eqref{full} implies that $i[K,L]=0$.

The operators 
\begin{equation*}
\A:=\hbox{$\frac12$}\(\Phi K+K\Phi\)\qquad\textrm{and}\qquad
\A':=\hbox{$\frac12$}\(\Phi L+L\Phi\)
\end{equation*}
are well-defined and symmetric on $D(\Phi)$. 

\begin{Lemma}\label{autoadjonction}
Let $\Phi$ be a function semi-adapted to the graph $(X,\sim)$.
\begin{enumerate}
\item[(a)] The operator $\A$ is essentially selfadjoint on $D(\Phi)$. The domain of
its closure $A$ is $D(A)=D(\Phi K)=\{f\in\H\:\!:\:\!\Phi Kf\in\H\}$ and $A$ acts on $D(A)$ as
the operator $\Phi K-\frac i2L$.
\item[(b)] The operator $\A'$ is essentially selfadjoint on $D(\Phi)$. The domain of
its closure $A'$ is $D(A')=D(\Phi L)=\{f\in\H\:\!:\:\!\Phi Lf\in\H\}$.
\end{enumerate}
\end{Lemma}

\begin{proof}
One just has to reproduce the proof of \cite[Lemma 3.1]{Georgescu/Golenia}, replacing their
couple $(N,S)$ by $(\Phi,K)$ for the point (a) and by $(\Phi,L)$ for the point (b).
\end{proof}

In the next lemma we collect some results on commutators with $A$ or $A'$.

\begin{Lemma}\label{Putnam}
Let $\Phi$ be a function semi-adapted to the graph $(X,\sim)$.
\begin{enumerate}
\item[(a)] The quadratic form $D(A)\ni f\mapsto i\<H f,Af\>-i\<Af,H f\>$ extends uniquely to
the bounded form defined by the operator $K^2$.
\item[(b)] The quadratic form $D(A)\ni f\mapsto i\<K^2f,Af\>-i\<Af,K^2f\>$ extends uniquely
to the bounded form defined by the operator $KLK+\hbox{$\frac12$}\(K^2L + LK^2\)$ (which
reduces to $2KLK$ if $\Phi$ is adapted).
\item[(c)] If $\Phi$ is adapted, the quadratic form
$D(A')\ni f\mapsto i\<K f,A'f\>-i\<A'f,K f\>$ extends uniquely to the bounded form defined by
the operator $L^2$.
\end{enumerate}
\end{Lemma}

The proof is straightforward. Computations may be performed on the core $D(\Phi)$. 
These results imply that $H\in C^1(A;\H)$, $K^2\in C^1(A;\H)$ and (when $\Phi$ is adapted)
$K\in C^1(A';\H)$.

Using the results of Lemma \ref{Putnam} we shall now establish a relation between the kernels
of the operators $H$, $K$ and $L$. For any selfadjoint operator $T$ in the Hilbert space $\H$
we write $\H_{\rm p}(T)$ for the closed subspace of $\H$ spanned by the eigenvectors of $T$.

\begin{Lemma}\label{lemeq}
For a function $\Phi$ semi-adapted to the graph $(X,\sim)$ one has
\begin{equation*}
\ker(H)\subset\H_{\rm p}(H)\subset\ker(K)\subset\H_{\rm p}(K).
\end{equation*}
If $\Phi$ is adapted, one also has 
\begin{equation*}
\H_{\rm p}(K)\subset\ker(L)\subset\H_{\rm p}(L).
\end{equation*}
\end{Lemma}

\begin{proof}
Let $f$ be an eigenvector of $H$. Due to the Virial Theorem \cite[Proposition 7.2.10]{ABG} and the
fact that $H$ belongs to $C^1(A;\H)$, one has $\<f,i[H,A]f\>=0$. It follows then by Lemma
\ref{Putnam}.(a) that $0=\<f,K^2f\>=\|Kf\|^2$, \ie $f\in\ker(K)$. The inclusion
$\H_{\rm p}(H)\subset \ker(K)$ follows. Similarly, by using $A'$ instead of $A$ and Lemma
\ref{Putnam}.(c) one gets the inclusion $\H_{\rm p}(K) \subset \ker(L)$ and the lemma is
proved.
\end{proof}

We are finally in a position to prove all the statements of Section \ref{sec3}.

\begin{proof}[Proof of Theorem \ref{unu}]
Since $H$ and $K$ are commuting bounded selfadjoint operators, the invariance of $\K$ and
$\G$ under $H$ and $K$ is obvious. Let us recall that $H_0$ and $K_0$ denote, respectively, 
the restrictions of the operators $H$ and $K$ to the subspace $\G$. By Lemma \ref{lemeq} one
has $\H_{\rm p}(H)\subset\K$, thus $H_0$ has no point spectrum.
\end{proof}

\begin{Lemma}\label{red}
If $\Phi$ is adapted to the graph $(X,\sim)$, then the decomposition $\H =\K\oplus\G$
reduces the operator $A$. The restriction of $A$ to $\G$ defines a selfadjoint operator
denoted by $A_0$.
\end{Lemma}

\begin{proof}
We already know that on $D(A)=D(\Phi K)$ one has $A=\Phi K-\frac{i}{2}L$. By using Lemma
\ref{lemeq} it follows that $\K\subset\ker A\subset D(A)$. Then one trivially checks that (i)
$A\[\K\cap D(A)\]\subset \K$, (ii) $A\[\G\cap D(A)\]\subset \G$ and
(iii) $D(A)=\[\K\cap D(A)\]+\[\G\cap D(A)\]$, which means that $A$ is reduced by the
decomposition $\H =\K\oplus\G$. Thus by \cite[Theorem 7.28]{Weidmann80} the restriction of $A$ 
to $D(A_0)\equiv D(A)\cap\G$ is selfadjoint in $\G$. 
\end{proof}

\begin{proof}[Proof of Theorem \ref{doi}]
We shall prove that the method of the weakly conjugate operator, presented in Section 
\ref{sec1}, applies to the operators $H_0$ and $A_0$ in the Hilbert space $\G$.

(i) Lemma \ref{Putnam}.(a) implies that $i(H_0 A_0-A_0 H_0)$ is equal in the form sense to
$K_0^2$ on $D(A_0)\equiv D(A)\cap\G$. Therefore the corresponding quadratic form extends
uniquely to the bounded form defined by the operator $K_0^2$. This implies that $H_0$ belongs
to $C^1(A_0;\G)$.

(ii) Since $B_0:=i[H_0,A_0]\equiv K_0^2>0$ in $\G$, the operator $A_0$ is weakly conjugate
to $H_0$. So we define the space $\BB$ as the completion of $\G$ with respect to the norm
$\|f\|_\BB:=\<f,B_0f\>^{1/2}$. The adjoint space of $\BB$ is denoted by $\BB^*$ and can be
identified with the completion of $B_0\G$ with respect to the norm
$\|f\|_{\BB^*}:=\<f,B_0^{-1}f\>^{1/2}$. It can also be expressed as the closure of the
subspace $K\H=K_0\G$ with respect to the same norm $\|f\|_{\BB^*}=\big\||K_0|^{-1}f\big\|$.
Due to Lemma \ref{Putnam}.(b) the quadratic form
$D(A_0)\ni f\mapsto i\<B_0 fA_0 f\>-i\<A_0 f,B_0 f\>$ extends uniquely to the bounded form
defined by the operator $2K_0L_0K_0$, where $L_0$ is the restriction of $L$ to $\G$. We
write $i[B_0,A_0]$ for this extension.

(iii) We check now that $\{W_t\}_{t\in\R}$ extends to a $C_0$-group in $\BB$. This easily
reduces to proving that for any $t\in\mathbb R$ there exists a constant $\textsc c(t)$ such
that $\|W_t f\|_\BB\leq\textsc c(t)\|f\|_\BB$ for all $f\in\mathcal D(A_0)$. Due to point
(ii) one has for each $f\in\mathcal D(A_0)$
\begin{equation*}
\left\|W_t f\right\|^2_\BB=\<f,B_0f\>+\int_0^t\d\tau\<W_\tau f,i[B_0,A_0]W_\tau f\>
\leq\|f\|^2_\BB+2\|L_0\|\int_0^{|t|}\d\tau\left\|W_\tau f\right\|_\BB^2.
\end{equation*}
Since $\G\hookrightarrow\BB$, the function $(0,|t|)\ni\tau\mapsto\|W_\tau f\|_\BB^2\in\R$ is
bounded. Thus we get the inequality $\|W_t f\|_\BB\leq\e^{|t|\|L_0\|}\|f\|_\BB$ by using a
simple form of the Gronwall Lemma. Therefore $\{W_t\}_{t \in \R}$ extends to a $C_0$-group
in $\BB$, and by duality $\{W_t\}_{t\in\R}$ also defines a $C_0$-group in $\BB^*$. It follows
immediately that the quadratic form $i[B_0,A_0]$ defines an element of $\B(\BB,\BB^*)$. This
concludes the proof of the fact that $B_0$ extends to an element of $C^1(A_0;\BB,\BB^*)$.
Thus all hypotheses of Theorem \ref{thmabstract} are satisfied and the limiting absorption
principle (\ref{liap}) holds for $H_0$, with $\EE$ given by
$\big(D(A_0,\BB^*),\BB^*\big)_{1/2,1}$.

(iv) \emph{A fortiori} the limiting absorption principle holds in the space $D(A_0, \BB^*)$ 
endowed with its graph norm. Let us show that the space $\FF$ introduced in Section \ref{sec3} 
is even smaller, with a stronger topology. We recall that for $f \in D(A_0,\BB^*)=
\{f\in D(A_0)\cap\BB^*\:\!:\:\!A_0f\in\BB^*\}$ (\cf \cite[Eq. 6.3.3]{ABG}) one has
\begin{equation*}
\|f\|_{D(A_0,\BB^*)}^2=\|f\|_{\BB^*}^2+\|A_0f\|_{\BB^*}^2=\big\||K_0|^{-1}f\big\|^2
+\big\||K_0|^{-1}A_0f\big\|^2.
\end{equation*}
We first prove that $K\H\cap D(\Phi)$ is dense in $\G$ and that
$K\H\cap D(\Phi)\subset D(A_0,\BB^*)$. For the density it is enough to observe that
$KD(\Phi)\subset K\H\cap D(\Phi)$ and that $KD(\Phi)$ is dense in $\G=\overline{K\H}$ since
$D(\Phi)$ is dense in $\H$ and $K$ is bounded. For the second statement, since $K\H=K_0\G$,
any $f$ in $K\H\cap D(\Phi)$ belongs to $\BB^*$ and to $D(A_0)=D(\Phi K)\cap\G$. Furthermore,
since $[K,L]=0$, we have $A_0f=K\Phi f+\frac{i}{2}Lf\in K\H\subset\BB^*$. This finishes to
prove that $K\H\cap D(\Phi) \subset D(A_0,\BB^*)$. We observe now that for $f$ in
$K\H\cap D(\Phi)$ one has 
\begin{equation*}
\big\||K_0|^{-1}A_0f\big\|=\big\||K_0|^{-1}\(K\Phi+\hbox{$\frac i2$}L\)f\big\|
\leq\|\Phi f\|+\hbox{$\frac 12$}\|L\|\big\||K_0|^{-1}f\big\|\leq\textsc c\:\!\|f\|_\FF
\end{equation*}
for some constant $\textsc c>0$ independent of $f$. It follows that
$\|f\|_{D(A_0,\BB^*)}\leq\textsc c'\|f\|_\FF$ for all $f\in K\H\cap D(\Phi)$ and a constant
$\textsc c'$ independent of $f$. Thus one has proved that $\FF\hookrightarrow\G$, and the
second continuous dense embedding $\G\hookrightarrow\FF^*$ is obtained by duality.
\end{proof}

\section{Admissible graphs}\label{secadmis}
\setcounter{equation}{0} 

In this section we put into evidence a class of graphs for which very explicit (and essentially
unique) adapted functions exist. For this class the spectral results are sharpened and
simplified.

Assume that the graph $(X,\sim)$ is connected and deduced from a directed graph, \ie some relation
$<$ is given on $X$ such that, for any $x,y\in X$, $x\sim y$ is equivalent to $x<y$ or $y<x$,
and one cannot have both $y<x$ and $x<y$. We also write $y>x$ for $x<y$, and note that $x<x$ is
forbidden.

Alternatively, one gets $(X,<)$ by decomposing for any $x\in X$ the set of neighbours of $x$
into a disjoint union, $N(x)=N^-(x)\sqcup N^+(x)$, taking care that $y\in N^-(x)$ if and only
if $x\in N^+(y)$. We call the elements of $N^-(x)$ \emph{the fathers} of $x$ and the elements 
of $N^+(x)$ \emph{the sons} of $x$, although this often leads to shocking situations.
Obviously, we set $x<y$ if and only if $x \in N^-(y)$, or equivalently, if and only if
$y\in N^+(x)$. When using drawings, one has to choose a direction (an arrow) for any edge. By
convention, we set $x\leftarrow y$ if $x<y$, \ie any arrow goes from a son to a father. When
directions have been fixed, we use the notation $(X,<)$ for the \emph{directed graph} and say
that $(X,<)$ is \emph{subjacent} to $(X,\sim)$.

Let $p=x_0x_1\ldots x_n$ be a path. Its \emph{index} is the difference between the number of
positively oriented edges and that of the negatively oriented ones, \ie
$\ind(p):=\#\{j\:\!:\:\!x_{j-1}<x_j\}-\#\{j\:\!:\:\! x_{j-1}>x_j\}$. The index is additive
under juxtaposition of paths: if $p=x_0x_1\ldots x_n$ and $q=y_0y_1\ldots y_m$ with
$x_n=y_0$, then the index of the path $pq:=x_0x_1\ldots x_{n-1}y_0y_1\ldots y_m$ is the sum of
the indices of the paths $p$ and $q$.

\begin{Definition}\label{admisibil}
A directed graph $(X,<)$ is called \emph{admissible} if
\begin{enumerate}
\item[(i)] it is \emph{univoque}, \ie any closed path in $X$ has index zero,
\item[(ii)] it is \emph{uniform}, \ie for any $x,y\in X$,
$\#[N^-(x)\cap N^-(y)]=\#[N^+(x)\cap N^+(y)]$.
\end{enumerate}
A graph $(X,\sim)$ is called \emph{admissible} if there exists an admissible directed graph
$(X,<)$ subjacent to $(X,\sim)$.
\end{Definition}

\begin{Definition}\label{posfun}
A \emph{position function} on a directed graph $(X,<)$ is a function $\Phi:X\to\Z$ satisfying
$\Phi(x)+1=\Phi(y)$ if $x<y$. 
\end{Definition}

We give now some properties of the position function.

\begin{Lemma}\label{propr}
\begin{enumerate}
\item[(a)] A directed graph $(X,<)$ is univoque if and only if it admits a position
function.
\item[(b)] Any position function on an admissible graph $(X,\sim)$ is surjective.
\item[(c)] A position function on a directed graph $(X,<)$ is unique up to a constant.
\end{enumerate}
\end{Lemma}

\begin{proof}
(a) Let $\Phi$ be a position function on $(X,<)$ and $p$ a path from $x$ to $y$. Then
$\ind(p)=\Phi(y)-\Phi(x)$. Thus $\ind(p)=0$ for any closed path. Conversely, assume univocity.
It is equivalent to the fact that, for any $x,y\in X$, each path from $x$ to $y$ has the same
index. Fix $z_0\in X$ and for any $z\in X$ set $\Phi(z):=\ind(p)$ for some path
$p=z_0z_1\ldots z$. Then $\Phi(z)$ does not depend on the choice of $p$ and is clearly a
position function.

(b) Since $\#N^-(x)=\#N^+(x)$ for any $x\in X$, it follows that each point of $X$ belongs to
a path which can be extended indefinitely in both directions.

(c) If $\Phi_1$ and $\Phi_2$ are two position functions and $p$ is a path from $x$ to $y$
(which exists since $X$ is connected), then $\Phi_1(y)-\Phi_1(x)=\ind(p)=\Phi_2(y)-\Phi_2(x)$,
thus $\Phi_1(y)-\Phi_2(y)=\Phi_1(x)-\Phi_2(x)$.
\end{proof}

Let us note that any univoque directed graph is bipartite, \ie it can be decomposed into two
disjoint subsets $X_1$, $X_2$ such that the edges connect only couples of the form
$(x_1,x_2)\in X_1\times X_2$. This is achieved simply by setting $X_1=\Phi^{-1}(2\Z+1)$ and
$X_2=\Phi^{-1}(2\Z)$. It follows then by \cite[Corollary 4.9]{Mohar/Woess} that the spectrum of
$H$ is symmetric with respect to the origin.

We are now in a position to prove Theorem \ref{first}.

\begin{proof}[Proof of Theorem \ref{first}]
We first show that for an admissible graph, any position function is adapted. Condition (i)
from Definition \ref{adaptat} is obvious. In the two remaining conditions of Definition
\ref{adaptat} one can decompose the sums over $N(x)\cap N(y)$ as sums over the four disjoint 
sets $N^-(x)\cap N^-(y)$, $N^+(x)\cap N^+(y)$, $N^-(x)\cap N^+(y)$ and $N^+(x)\cap N^-(y)$. 
In the last two cases the sums are zero and in the other two cases the sums give together 
$2\big(\#[N^+(x)\cap N^+(y)]-\#[N^-(x)\cap N^-(y)]\big)$, which is also zero by the
uniformity of the graph. 

Therefore Theorem \ref{doi} can be applied. If $\Phi$ is a position function, one has
$\Phi(y)-\Phi(x)=\pm1$  if $x\sim y$ and thus $L=-H$. Consequently, Lemma \ref{lemeq} gives
the equalities
\begin{equation*}
\H_{\rm p}(H)=\ker(K)=\H_{\rm p}(K)=\ker(H)=\left\{f\in\H\:\!:\:\!\textstyle\sum_{y>x}f(y)=0
=\sum_{y<x}f(y)~\textrm{for each}~x\in X\right\}
\end{equation*}
which complete the proof.
\end{proof}

Note that even when $\K\ne\{0\}$ the singular continuous spectrum of $H$ is empty. Indeed,
in the canonical decomposition $\H=\H_{\rm p}(H)\oplus\H_{\rm ac}(H)\oplus\H_{\rm sc}(H)$, 
$\H_{\rm p}(H)$ is identified with $\K$, $\H_{\rm ac}(H)$ with $\G$, and $\H_{\rm sc}(H)$ 
is thus trivial. Furthermore, a look at the proof above shows that the results of Theorem
\ref{first} hold in fact for any graph with an adapted function $\Phi$ satisfying
$\Phi(y)-\Phi(x)=\pm1$ if $x\sim y$. We decided to insist on the particular case of
admissible graphs because admissibility can be checked straightforwardly by inspecting the
subjacent directed graph; in case of successful verification the function $\Phi$ is generated
automatically.

\begin{Remark}
For a directed graph $(X,<)$, define $(Uf)(x):=\sum_{y<x}f(y)$ for each $f\in\H$ and
$x\in X$. The operator $U$ is bounded and its adjoint is given by
$(U^*f)(x)=\sum_{y>x}f(y)$. One has $H=2\re U$ and $K=2\im U$. Uniformity of $(X,<)$ is
equivalent to the normality of $\,U$, thus to the fact that $H$ and $K$ commute. In
\cite{Georgescu/Golenia} it is shown that the adjacency operator of a homogeneous rooted
tree can be written as $H=2\re U$ for $U$ a completely non unitary isometry (\ie an
isometry such that $U^{*n}\to0$ strongly). This fact is used to prove the existence of
an operator $N$ (called \emph{number operator}) sastisfying $UNU^*=N-1$. $N$ is used to
construct an operator $A=N(\im U)+(\im U)N$, which is conjugate (in the sense of Mourre
theory) to $H$ and to some classes of perturbations of $H$. Note that $N$ is not a
multiplication operator. It would be interesting to find an approach unifying the present
study with the work \cite{Georgescu/Golenia}.
\end{Remark}

One can show that finite cartesian products of admissible directed graphs are admissible.
Indeed uniformity follows rather easily from the definitions and, if $\Phi_j$ is the
position function for $(X_j,<_j)$, then $\Phi$ defined by
$\Phi(x_1,\ldots,x_n):=\sum_{j=1}^n\Phi_j(x_j)$ is the natural position function for the
cartesian product $\prod_j(X_j,<_j)$. As an example, $\Z^n$ is admissible, since $\Z$ is
obviously admissible. We shall not give details here since these are simple facts, 
largely covered by Section \ref{Dp}.

\section{Examples}\label{someother}
\setcounter{equation}{0} 

We present some examples of graphs (admissible or not) with an adapted function which can be
easily drawn in the plane. Although some of them might be subject to other treatments, we would
like to stress the relative ease and unity of our approach, which also furnishes boundary
estimates for the resolvent and applies to some classes of perturbations. In many situations we
will be able to determine the kernel $\K$ of the operator $K$ explicitly. In the case
$\K=\{0\}$ the graph is said to be \emph{injective}. For admissible graphs, we recall that
$\ker(K)=\ker(H)$ coincides with the singular subspace of $H$ and that it is given by Formula
\eqref{ma_belle_courgette}.

The directed graph $X$ of Figure \ref{graph1} is admissible, non-regular and not injective.
\begin{figure}[htbp]
\begin{center}
\input{graph1.pstex_t}
\caption{\textsf{\footnotesize Example of an admissible, non-injective directed graph}}
\label{graph1}
\vspace{-10pt}
\end{center}
\end{figure}
Indeed, $\K$ is composed of all $f\in\ell^2(X)$ taking the value $0$ on the middle row and
opposite values on the other two rows. 
\begin{figure}[htbp]
\begin{center}
\input{graph2.pstex_t}
\caption{\textsf{\footnotesize Example of an admissible, non-injective directed graph}}
\label{graph2}
\vspace{-10pt}
\end{center}
\end{figure}
Other examples of graphs of this type are available (see for example Figure \ref{graph2}).

One can sometimes construct admissible graphs $X$ by juxtaposing admissible graphs in some
coherent manner. For instance the directed graph of Figure \ref{graph3} is admissible and
injective, so that its adjacency operator is purely absolutely continuous.
\begin{figure}[htbp]
\begin{center}
\input{graph3.pstex_t}
\caption{\textsf{\footnotesize Example of an admissible, injective directed graph}}
\label{graph3}
\vspace{-10pt}
\end{center}
\end{figure}
Writting the condition $\sum_{w<x}f(w)=0$ for $f\in\ell^2(X)$ and $x$ as in Figure
\ref{graph3}, one gets $f(z)=0$. But one has also $f(z)+f(z')=0$ due to the same condition
for the vertex $y$. Thus $f(z')=0$, and the graph is injective since the same argument
holds for each vertex of $X$. Extension of the graph in both vertical directions induces the
standard Cayley graph of $\Z^2$, which is clearly admissible and injective. If we extend the
graph only downwards, then we obtain the subgraph $\{(x_1,x_2)\in\Z^2\:\!:\:\!x_1<x_2\}$,
which is also admissible and injective (see the final paragraph of Section \ref{secadmis}).

\begin{figure}[htbp]
\begin{center}
\input{graph4.pstex_t}
\caption{\textsf{\footnotesize Examples of admissible, injective directed graphs}}
\label{graph4}
\vspace{-10pt}
\end{center}
\end{figure}

The directed graph of Figure \ref{graph4}.(a) is admissible and injective, so its adjacency
operator has no singular continuous spectrum and no point spectrum. One shows easily that
admissibility and injectivity are preserved under a finite or infinite number of vertical
juxtapositions of the graph with itself (see Figure \ref{graph4}.(b)). On the other hand, if
one puts Figure \ref{graph4}.(a) on top of itself, deletes all the arrows belonging to the
middle row as well as the vertices left unconnected, one gets an admissible, non-injective
directed graph.

\begin{figure}[htbp]
\begin{center}
\input{graph5.pstex_t}
\caption{\textsf{\footnotesize Example of an admissible, non-injective directed graph}}
\label{graph5}
\vspace{-10pt}
\end{center}
\end{figure}

The directed graph of Figure \ref{graph5} is admissible, regular but not injective. The
graph deduced from it is the Cayley graph of $\Z\times\Z_2$, with generating system
$\{(\pm1,1),(\pm1,-1)\}$, without being a cartesian product. The elements of $\K$ are all
$\ell^2$-functions which are anti-symmetric with respect to a vertical flip. If two copies
of this graph are juxtaposed vertically, the resulting graph is still admissible, but also
regular and injective. If one deletes some chosen arrows in the resulting graph, one obtains 
a nice admissible, non-injective graph with vertices of degree $2$, $4$ and $6$ 
(see Figure \ref{graph9}).

\begin{figure}[htbp]
\begin{center}
\input{graph9.pstex_t}
\caption{\textsf{\footnotesize Example of an admissible, non-injective directed graph}}
\label{graph9}
\vspace{-10pt}
\end{center}
\end{figure}

Admissible graphs are of a very restricted type. For instance closed paths of odd length
and vertices of odd degree are forbidden. We give now a few more examples of graphs, for
which non-constant adapted functions $\Phi$ exist. At each vertex, the indicated number corresponds
to the value of $\Phi$.

\begin{figure}[htbp]
\begin{center}
\input{graph8.pstex_t}
\caption{\textsf{\footnotesize Example of a non-admissible, adapted, injective graph}}
\label{graph8}
\vspace{-10pt}
\end{center}
\end{figure}

Easy computations show that the function $\Phi$ associated with the non-admissible regular graph 
of Figure \ref{graph8} is adapted. Furthermore, this graph is injective. This is not
unexpected, since it is a very simple Cayley graph of the abelian group $\Z\times\Z_2$.
Deleting steps in this ladder graph leads generically to singular continuous spectrum as
pointed out in \cite{Simon96}.

\begin{figure}[htbp]
\begin{center}
\input{graph6.pstex_t}
\caption{\textsf{\footnotesize Example of a non-admissible, adapted, non-injective graph}}
\label{graph6}
\vspace{-10pt}
\end{center}
\end{figure}

\begin{figure}[htbp]
\begin{center}
\input{graph7.pstex_t}
\caption{\textsf{\footnotesize Example of a non-admissible, adapted graph}}
\label{graph7}
\vspace{-10pt}
\end{center}
\end{figure}

The function $\Phi$ indicated for the non-admissible regular graph of Figure \ref{graph6} is adapted. 
One shows easily that the space $\K$ coincides with the eigenspace of the adjacency operator $H$ associated
with the eigenvalue $-1$. The rest of the spectrum is purely absolutely continuous.
The function $\Phi$ of the non-admissible non-regular graph of Figure \ref{graph7} is adapted. However, we
believe that this graph is not injective. More graphs with an adapted function will be indicated 
in the next section.

\section{$\boldsymbol D$-Products}\label{Dp}
\setcounter{equation}{0} 

We recall now some properties of adjacency operators on the class of \emph{$D$-products}. We
call $D$-product what is referred as \emph{non-complete extended $p$-sum with basis $D$} in
\cite{Mohar/Woess}.

Consider a family $\{(X_j,\sim_j)\}_{j=1}^n$ of simple graphs, which are all infinite countable
and uniformly locally finite. Let $D$ be a subset of $\{0,1\}^n$ not containing $(0,0,\dots,0)$.
Then we endow the product $X:=\prod_{j=1}^nX_j$ with the following ($D$-product) graph
structure: if $x,y\in X$ then $x\sim y$ if and only if there exists $d\in D$ such that
$x_j\sim_jy_j$ if $d_j=1$ and $x_j=y_j$ if $d_j=0$. The resulting graph $(X,\sim)$ is again
simple, infinite countable and uniformly locally finite. Note that the tensor product as well
as the cartesian product are special cases of $D$-product. We shall not assume $(X_j,\sim_j)$
connected and even if we did, the $D$-product could fail to be so. 

It is easy to see that the adjacency operator $H$ of the $D$-product $(X,\sim)$ may be written as 
\begin{equation}\label{D-prod}
H=\sum_{d\in D}H_1^{d_1}\otimes\dots\otimes H_n^{d_n},
\end{equation}
where $H_j$ is the adjacency operator of $(X_j,\sim_j)$, $H_j^0=1$ and $H_j^1=H_j$. The operator $H$ 
acts in the Hilbert space $\ell^2(X)\simeq\bigotimes_{j=1}^n\ell^2(X_j)$.

\begin{Proposition}\label{D-P}
For each $j\in\{1,\dots,n\}$, let $\Phi_j$ be a function adapted to the graph $(X_j,\sim_j)$ and 
$c_j\in\R$. Then $\Phi_c:X\to\R$, $(x_1,\dots,x_n)\mapsto\sum_{j=1}^nc_j\Phi_j(x_j)$ is a function adapted to 
$(X,\sim)$.
\end{Proposition}

\begin{proof}
Rather straightforward calculations show that $\Phi_c$ satisfies (\ref{semi}) and (\ref{full}).
It is simpler to indicate a simpler operatorial proof:

Define $K_j:=i[H_j,\Phi_j]$ and $L_j:=i[K_j,\Phi_j]$ in $\ell^2(X_j)$. Since $\Phi_j$ is 
adapted the three operators $H_j, K_j$ and $L_j$ commute (use the Jacobi identity for the triple 
$H_j, K_j$ and $\Phi_j$). Since the multiplication operator $\Phi_c$ can be written in $\bigotimes_j 
\ell^2(X_j)$ as $\Phi_c=\sum_{j=1}^n c_j\,1\otimes\dots\otimes\Phi_j\otimes\dots\otimes 1$,
where $\Phi_j$ stands on the $j$'th position, one has 
\begin{equation*}
K:=i[H,\Phi_c]=\sum_{d\in D}\sum_j\,c_j\,H_1^{d_1}\otimes\dots\otimes K_j(d_j)\otimes
\dots\otimes H_n^{d_n},
\end{equation*}
where $K_j(d_j)$ stands on the $j$'th position and is equal to $K_j$ if $d_j=1$ and to $0$ if 
$d_j=0$. Analogously one has 
\begin{eqnarray*}
L:=i[K,\Phi_c]\!\!&=&\!\!\sum_{d\in D}\sum_{j\ne k}\,c_jc_k\,H_1^{d_1}\otimes\dots\otimes K_j(d_j)
\otimes\dots\otimes K_k(d_k)\otimes\dots\otimes H_n^{d_n} \\
&&+\sum_{d\in D}\sum_j\,c_j^2\,H_1^{d_1}\otimes\dots\otimes L_j(d_j)\otimes\dots\otimes H_n^{d_n},
\end{eqnarray*}
where $L_j(d_j)$ is equal to $L_j$ if $d_j=1$ and to $0$ if $d_j=0$. It is clear that $i[H,K]=0=i[K,L]$, 
which is equivalent to the statement of the proposition.
\end{proof}

Notice that we could very well have no valuable information on some of the graphs $X_j$ and take 
$\Phi_j=0$. As soon as $\Phi_c$ is not a constant, the space $\G$ on which we have a purely absolutely 
continuous restricted operator is non-trivial. So one can perform various $D$-products, including factors 
for which an adapted function has already been shown to exist (as those in the preceding section). 
But it is not clear how the space $\K=\ker(K)$ could be described in such a generality.

\section{The one-dimensional XY model}\label{xysection}
\setcounter{equation}{0} 

In the sequel we apply the theory of Section \ref{secadmis} to the Hamiltonian of the
one-dimensional XY model. We follow \cite{DMT} for the brief and rather formal presentation of
the model. Further details may be found in \cite{Streater67}, \cite{Streater74} and
\cite[Sec. 6.2.1]{Bratelli/Robinson2}.

We consider the one-dimensional lattice $\Z$ with a spin-$\frac12$ attached at each vertex. Let 
\begin{equation*} 
\F(\Z):=\{\alpha:\Z\to\{0,1\}\:\!:\:\!\supp(\alpha)\ \text{is finite}\}\:\!, 
\end{equation*} 
and write $\{e^0,e^1\}:=\{(0,1),(1,0)\}$ for the canonical basis of the (spin-$\frac12$) Hilbert
space $\mathbb C^2$. For any $\alpha\in\F(\Z)$ we denote by $e^\alpha$ the element
$\{e^{\alpha(x)}\}_{x\in \Z}$ of the direct product $\prod_{x\in \Z}\mathbb C^2_x$. We 
distinguish the vector $e^{\alpha_0}$, where $\alpha_0(x):=0$ for all $x\in \Z$. Each element 
$e^\alpha$ is interpreted as a state of the system of spins, and $e^{\alpha_0}$ as its
ground state with all spins pointing down. The Hilbert space $\M$ of the system (which is
spanned by the states with all but finitely many spins pointing down) is the ``incomplete
tensor product'' \cite[Sec. 2]{Streater67}, \cite[Sec. 2]{Streater74} 
\begin{equation*} 
\M:=\bigotimes_{x\in\Z}^{\alpha_0}\mathbb C^2_x\equiv\text{closed span} 
\left\{e^\alpha\:\!:\:\!\alpha\in\F(\Z)\right\}. 
\end{equation*} 
The dynamics of the spins is given by the nearest-neighbour XY Hamiltonian 
\begin{equation*}
M:=-\mbox{$\frac12$}\sum_{|x-y|=1}
\(\sigma_1^{(x)}\sigma_1^{(y)}+\sigma_2^{(x)}\sigma_2^{(y)}\).
\end{equation*}
The operator $\sigma_j^{(x)}$ acts in $\M$ as the identity operator on each factor $\C^2_y$,
except on the component $\C^2_x$ where it acts as the Pauli matrix $\sigma_j$.
To go further on, we need to introduce a new type of directed graphs.

\begin{Definition}\label{directed_FN}
Let $(X,<)$ be a directed graph. For $N\in\N$, we set
$\F_N(X):=\{\alpha:X\to\{0,1\}\:\!:\:\!\#\supp(\alpha) = N\}$ and endow it with the 
natural directed graph structure defined as follows: if $\alpha,\beta\in\F_N(X)$
then $\alpha<\beta$ if and only if there exist $x\in\supp(\alpha)$, $y\in\supp(\beta)$ 
such that $x<y$ and $\supp(\alpha)\setminus\{x\}=\supp(\beta)\setminus\{y\}$.
\end{Definition}

From now on, we shall no longer make any distinction between an element $\alpha\in\F_N(X)$
and its support, which is a subset of $X$ with $N$ elements. We recall from \cite[Sec. 2]{DMT} 
that $M$ is unitarily equivalent to a direct sum $\bigoplus_{N\in\N}H_N$, where $H_N$ is the 
selfadjoint operator in $\H_N:=\ell^2\[\F_N(\Z)\]$ acting as 
\begin{equation*}
(H_Nf)(\alpha)=-2\sum_{\beta\sim\alpha}f(\beta),\quad f\in\H_N,~\alpha\in\F_N(\Z).
\end{equation*}

Thus the spectral analysis of $M$ reduces to determining the nature of the spectrum of the
adjacency operators on $\H_N$. Moreover the graph $\(\F_N(\Z),\sim\)$ deduced from
$\(\F_N(\Z),<\)$ satisfies

\begin{Lemma}\label{efen}
$\(\F_N(\Z),\sim\)$ is an admissible graph. 
\end{Lemma}

\begin{proof}
Due to Definition \ref{admisibil} one simply has to prove that $\(\F_N(\Z),<\)$ is
admissible. In point (i) we show that $\(\F_N(\Z),<\)$ is uniform. In point (ii) we give
the (natural) position function for $\(\F_N(\Z),<\)$.

(i) Given $\alpha\in\F_N(\Z)$ and $x\in\supp(\alpha)$,
$y\notin\supp(\alpha)$, we write $\alpha^y_x$ for the function of $\F_N(\Z)$ such that
$\supp(\alpha^y_x)=\supp(\alpha)\sqcup\{y\}\setminus\{x\}$.
Thus one has
\begin{equation*}
N^-(\alpha)\cap N^-(\beta)
=\left\{\gamma\:\!:\:\!\exists x\in\alpha,\,x-1\notin\alpha,~
\exists y\in\beta,\,y-1\notin\beta,~\gamma=\alpha^{x-1}_x=\beta^{y-1}_y\right\}
\end{equation*}
and
\begin{equation*}
N^+(\alpha)\cap N^+(\beta)
=\left\{\gamma\:\!:\:\!\exists x\in\alpha,\,x+1\notin\alpha,~
\exists y\in\beta,\,y+1\notin\beta,~\gamma=\alpha^{x+1}_x=\beta^{y+1}_y\right\},
\end{equation*} 
the couples $(x,y)$ being unique for a given $\gamma$ in both cases.
Suppose there exist $x\in\alpha$, $y\in\beta$ such that $x-1\notin\alpha$,
$y-1\notin\beta$ and $\alpha^{x-1}_{x}=\beta^{y-1}_{y}$, so that
$\alpha^{x-1}_{x}\in\{N^-(\alpha)\cap N^-(\beta)\}$. If $x=y$, then $\alpha=\beta$, and
$\#N^-(\alpha)$, $\#N^+(\alpha)$ are both equal to the number of connected components of
$\alpha$. If $x \neq y$, then one has $x-1\in\beta$, $x\notin\beta$, $y-1\in\alpha$,
$y\notin\alpha$ together with the equality $\alpha^{y}_{y-1}=\beta^{x}_{x-1}$. Therefore
$\alpha^{y}_{y-1}\in\{N^+(\alpha)\cap N^+(\beta)\}$ and one has thus obtained a bijective
map from $N^-(\alpha)\cap N^-(\beta)$ to $N^+(\alpha)\cap N^+(\beta)$.

(ii) If $\Phi_\Z$ is a position function for $\Z$ (for instance $\Phi_\Z(x)=x$), it is
easily checked that $\Phi$ defined by $\Phi(\alpha):=\sum_{x\in\alpha}\Phi_\Z(x)$ is a
position function for $\F_N(\Z)$.
\end{proof}

\begin{Remark}
One could presume that $\(\F_N(\Z^2),<\)$ is also an admissible
directed graph. But this is wrong, as it can be seen from the following example. For $N=2$,
consider $\alpha:=\{(1,0),(1,1)\}$ and $\beta:=\{(0,1),(1,1)\}$. It can be easily checked
that $N^-(\alpha)\cap N^-(\beta)=\big\{\{(0,0),(1,1)\},\{(1,0),(0,1)\}\big\}$, whereas
$N^+(\alpha)\cap N^+(\beta)=\varnothing$. This contradicts the uniformity hypothesis.
\end{Remark}

As a corollary of Theorem \ref{first} and of the admissibility of $\(\F_N(\Z),\sim\)$, one
obtains:

\begin{Corollary}\label{presque}
The spectrum of $M$ is purely absolutely continuous, except maybe at the origin.
\end{Corollary}

\begin{Remark}\label{pourquoipresque}
We would obtain that the spectrum of $M$ is purely absolutely continuous
if we could show that $\ker(H_N)=\{0\}$ for any $N$. Unfortunately
we have been able to obtain such a statement only for $N=1,2,3$ and $4$.
Our proof consists in showing that if there exists $f\in\ker(H_N)$ such
that $f(\alpha)\neq0$ for some $\alpha\in\F_N(\Z)$, then there exists an
infinite number of elements $\alpha'\in \F_N(\Z)$ such that
$f(\alpha')=f(\alpha)$, which contradicts the requirement
$f\in\ell^2[\F_N(\Z)]$. Even if we did not succeed in extending such an
argument for $N>4$, we do believe that the kernel of $H_N$ is trivial for
any $N$.
\end{Remark}

\section*{Acknowledgements}

S.R. is supported by the european network: Quantum Spaces - Noncommutative Geometry. R.T.d.A. is
supported by the Swiss National Science Foundation. This work was initiated while M.M. was
visiting the universities of Lyon and Geneva. He would like to thank Professor Johannes Kellendonk, 
the members of the DPT (Geneva) and especially Professor Werner Amrein for their kind hospitality.


\end{document}